\documentclass[aps,prl,twocolumn,superscriptaddress]{revtex4-1}

\usepackage{graphicx}
\usepackage{MnSymbol}
\usepackage{todonotes}
\usepackage{hyperref}
\usepackage{soul}
\begin{document}

\title{Persistence of magnetic excitations in La$_{2-x}$Sr$_x$CuO$_4$ from the undoped insulator to the heavily overdoped non-superconducting metal}

\author{M. P. M. Dean}
\email{mdean@bnl.gov}
\affiliation{Department of Condensed Matter Physics and Materials Science,Brookhaven National Laboratory, Upton, New York 11973, USA}

\author{G. Dellea}
\affiliation{Dipartimento di Fisica, Politecnico di Milano, Piazza Leonardo da Vinci 32, I-20133 Milano, Italy}

\author{R. S. Springell}
\affiliation{Royal Commission for the Exhibition of 1851 Research Fellow, Interface Analysis Centre, University of Bristol, Bristol BS2 8BS, UK}

\author{F. Yakhou-Harris}
\author{K. Kummer}
\author{N. B. Brookes}
\affiliation{European Synchrotron Radiation Facility (ESRF), BP 220, F-38043 Grenoble Cedex, France}

\author{X. Liu}
\affiliation{Beijing National Laboratory for Condensed Matter Physics, and Institute of Physics, Chinese Academy of Sciences, Beijing 100190, China}
\affiliation{Department of Condensed Matter Physics and Materials Science,Brookhaven National Laboratory, Upton, New York 11973, USA}

\author{Y.-J. Sun}
\affiliation{Beijing National Laboratory for Condensed Matter Physics, and Institute of Physics, Chinese Academy of Sciences, Beijing 100190, China}
\affiliation{Department of Condensed Matter Physics and Materials Science,Brookhaven National Laboratory, Upton, New York 11973, USA}

\author{J. Strle}
\affiliation{Department for Complex Matter, Jo\v{z}ef Stefan Institute, Jamova 39, 1000 Ljubljana, Slovenia}
\affiliation{Department of Condensed Matter Physics and Materials Science,Brookhaven National Laboratory, Upton, New York 11973, USA}

\author{T. Schmitt}
\affiliation{Swiss Light Source, Paul Scherrer Institut, CH-5232 Villigen PSI, Switzerland}

\author{L. Braicovich}
\author{G. Ghiringhelli}
\affiliation{Dipartimento di Fisica, Politecnico di Milano, Piazza Leonardo da Vinci 32, I-20133 Milano, Italy}
\affiliation{CNR-SPIN, Consorzio Nazionale Interuniversitario per le Scienze Fisiche della Materia, Italy}

\author{I. Bo\v{z}ovi\'{c}}
\author{J. P. Hill}
\email{hill@bnl.gov}
\affiliation{Department of Condensed Matter Physics and Materials Science,Brookhaven National Laboratory, Upton, New York 11973, USA}

\def\mathbi#1{\ensuremath{\textbf{\em #1}}}
\def\Q{\ensuremath{\mathbi{Q}}}
\newcommand{\angstrom}{\mbox{\normalfont\AA}}
\date{\today}
%
%
%
%
\pacs{74.70.Xa,75.25.-j,71.70.Ej}
%
\maketitle

\textbf{
One of the most intensely studied scenarios of high-temperature superconductivity (HTS) postulates pairing by exchange of magnetic excitations \cite{Scalapino2012}. Indeed, such excitations have been observed up to around optimal doping in the cuprates \cite{Bourges2000, Tranquada2004, Hayden2004, Vignolle2007, Xu2009, LeTacon2011}. In the heavily overdoped regime, neutron scattering measurements indicate that magnetic excitations have effectively disappeared \cite{Wakimoto2004, Wakimoto2007, Fujita2012NS}, and this was argued to cause the demise of HTS with overdoping \cite{Wakimoto2004, Fujita2012NS, Scalapino2012}.
Here we use resonant inelastic x-ray scattering (RIXS), which is sensitive to complementary parts of reciprocal space, to measure the evolution of the magnetic excitations in La$_{2-x}$Sr$_x$CuO$_4$ across the entire phase diagram, from a strongly correlated insulator ($x=0$) to a non-superconducting metal ($x=0.40$). For $x=0$, well-defined magnon excitations are observed \cite{Coldea2001}. These magnons broaden with doping, but they persist with a similar dispersion and comparable intensity all the way to the non-superconducting, heavily overdoped metallic phase. The destruction of HTS with overdoping is therefore caused neither by the general disappearance nor by the overall softening of magnetic excitations; other factor(s), such as the redistribution of spectral weight, must be considered.
}

\begin{figure*}
\includegraphics{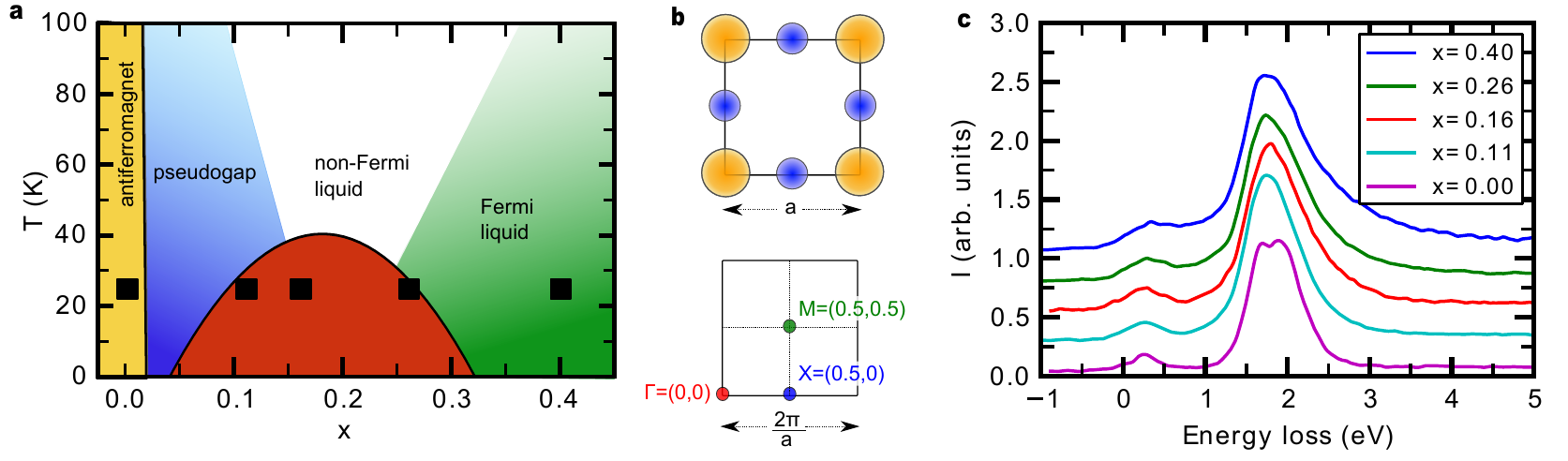} %
\caption{\textbf{La$_{2-x}$Sr$_x$CuO$_4$ diagrams and RIXS spectra}  (a) A schematic phase diagram for La$_{2-x}$Sr$_x$CuO$_4$ as a function of the doping level, $x$. The black lines defining the superconducting and antiferromagnetic states are based on data from Ref.~\cite{Keimer1992}. The doping levels of the samples studied here, and the temperature at which the measurements were made, are marked by black squares. (b) Top, the basic structural unit of the cuprates: a CuO$_2$ plaquette with Cu atoms in orange and O atoms in blue separated by distance $a\approx3.8$~\angstrom{}. Bottom, the reciprocal lattice with high symmetry points marked and labeled in reciprocal lattice units (r.\ l.\ u.).  (c)  The RIXS spectrum of La$_{2-x}$Sr$_x$CuO$_4$ for $x=0$, $x=0.11$, $x=0.16$, $x=0.26$, and $x=0.40$ at  $\Q = (0.36, 0)$. The peaks in the 1-3~eV energy window are the $dd$-excitations, while the low energy peak around 300~meV arises from magnetic scattering.}
\label{Fig1}
\end{figure*}

The undoped high-$T_c$ cuprates such as La$_2$CuO$_4$ are antiferromagnetic (N\'{e}el-ordered) insulators, with magnetic Bragg peaks and well-defined high-energy magnetic excitations termed magnons \cite{Coldea2001}. As shown in Fig.~\ref{Fig1}(a), doping rapidly destroys the N\'{e}el ordering, leading to the emergence of the pseudogap state and superconductivity. In the underdoped and optimally doped cuprates, superconductivity is accompanied by an ``hour-glass'' shaped dispersion of magnetic excitations around the scattering vector $\Q_{\text{AFM}}=(0.5,0.5)$ in Fig.~\ref{Fig1}(b) \cite{Arai1999,Bourges2000, Tranquada2004, Hayden2004, Vignolle2007, Xu2009}. In the lightly overdoped, but still superconducting regime, high energy magnetic excitations have been observed in La$_{1.78}$Sr$_{0.22}$CuO$_4$ \cite{Lipscombe2007} and YBa$_2$Cu$_3$O$_7$ \cite{LeTacon2011}. Far less work has been done on the magnetic excitations in the heavily overdoped region of the phase diagram. Neutron scattering studies of La$_{1.70}$Sr$_{0.30}$CuO$_4$ report that the \Q{}-integrated magnetic dynamic structure factor $S(\omega)$ is much reduced by $x=0.25$ and that magnetic excitations have effectively disappeared by $x=0.30$ \cite{Wakimoto2007}. This observation has been used in support of proposals that spin fluctuations mediate the electron pairing in high-$T_c$ superconductors \cite{Scalapino2012}. A necessary, although not sufficient, condition for such scenarios is that spin fluctuations persist across the superconducting portion of the phase diagram while retaining appreciable spectral weight. For this reason it was suggested that the destruction of HTS in the overdoped cuprates is due to the disappearance of magnetic excitations \cite{Wakimoto2004}.

Resonant inelastic x-ray scattering (RIXS) at the Cu $L_3$ edge has recently emerged as a new experimental method for measuring magnetic excitations in the cuprates \cite{Braicovich2009, Braicovich2010, LeTacon2011, Dean2012, Dean2013a, Zhou2013, Dean2013b}. RIXS is particularly well suited to measuring high energy magnetic excitations and requires only very small sample volumes \cite{Braicovich2009,Braicovich2010}. As explained by Le Tacon \emph{et al.}(Ref.~\cite{LeTacon2011}), this sets it apart from current neutron scattering experiments which require large single crystals of several cm$^3$ in volume, which are usually very difficult to synthesize, especially in the heavily overdoped region. We also note that Cu $L_3$ edge RIXS experiments focus on a complementary region of the Brillouin zone in Fig.~\ref{Fig1}(b) compared to most \Q{}-resolved neutron scattering experiments \cite{Ament2011}. RIXS typically measures from $(0,0)$ towards $(0.5,0)$; whereas neutron scattering experiments focus around $(0.5,0.5)$, where the magnetic excitations are strongest.

We synthesized La$_{2-x}$Sr$_x$CuO$_4$ films with $x=0$, $0.11$, $0.16$, $0.26$ and $0.40$ using molecular beam epitaxy (MBE). These films, unlike bulk samples, have atomically smooth surfaces (root mean square roughness, as measured by Atomic Force Microscopy, down to a few $\angstrom$), which reduces the diffuse elastic scattering contribution to the spectra  \cite{Ament2011}. We chose the doping levels to span the La$_{2-x}$Sr$_x$CuO$_4$ phase diagram as indicated by the solid black squares in Fig.~\ref{Fig1}(a).

RIXS spectra for these samples are shown in Fig.~\ref{Fig1}(c). The most intense feature corresponds to optically-forbidden $dd$ orbital excitations in which the valence band hole, primarily of Cu $d_{x^2-y^2}$ character, is promoted into  higher energy orbitals \cite{Ghiringhelli2004}. The intensity of these excitations can provide a reference to compare different RIXS spectra \cite{BraicovichPRB2010}. In the mid-infrared (MIR) energy scale (50-500~meV) single spin-flip excitations can be excited due to the spin-orbit coupling of the Cu $2p_{3/2}$ core hole \cite{Ament2009, Haverkort2010}. A broad, flat background of intensity arises from charge-transfer excitations of the Cu $d_{x^2-y^2}$ hole into the O $2p$ states. As $x$ increases the $dd$ excitations are seen to broaden, indicative of hybridization between the $d$-orbitals and the itinerant states of doped holes. This leads to a tail of intensity extending down into the MIR energy region.

\begin{figure*}
\includegraphics{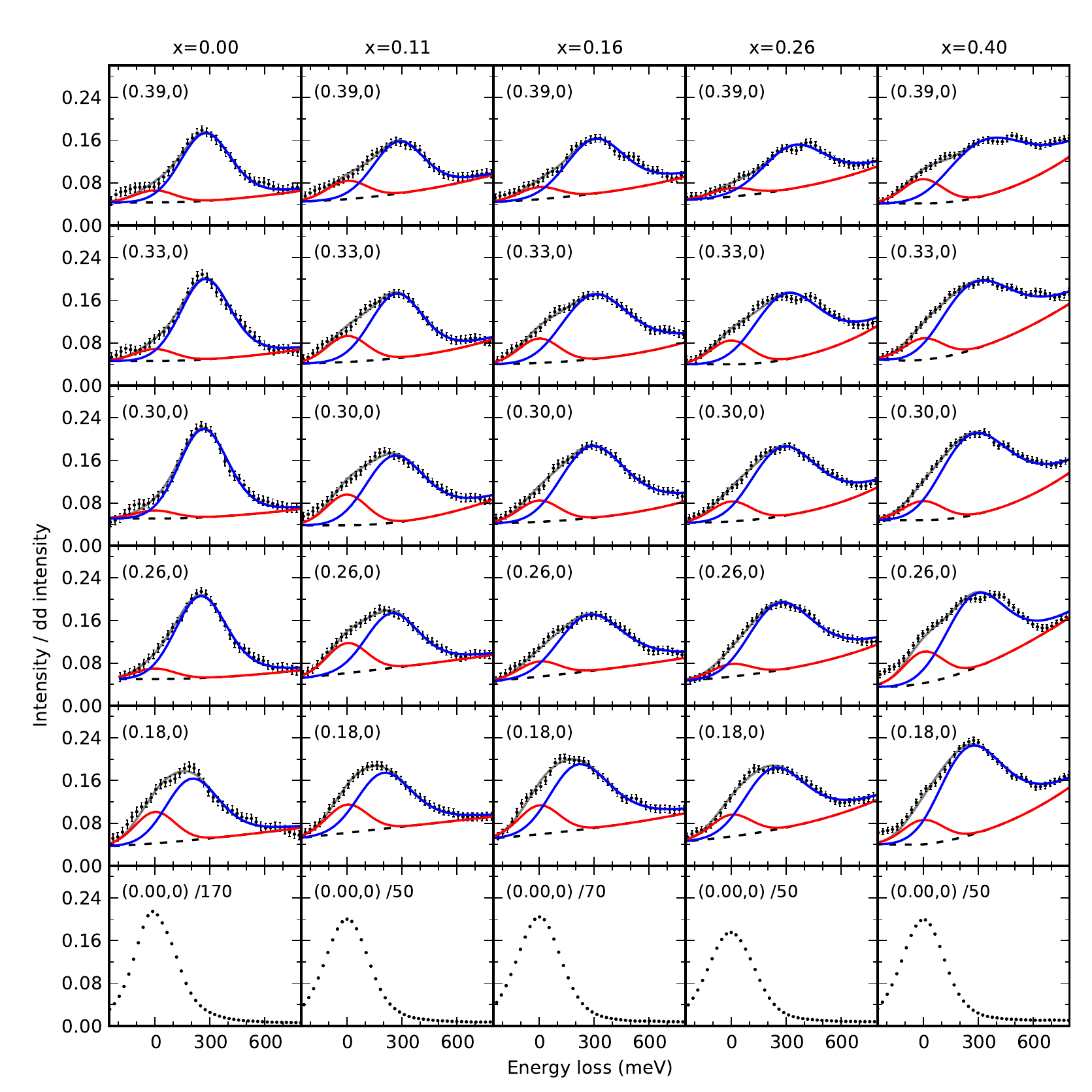} %
\caption{\textbf{La$_{2-x}$Sr$_x$CuO$_4$ magnetic excitation spectra.} The dispersion of the magnetic excitations in La$_{2-x}$Sr$_x$CuO$_4$ as function of \Q{} and $x$. The filled black circles represent the data and the solid grey line shows the results of the fitting, which is the sum of an elastic line (red), an anti-symmetrized Lorentzian capturing the magnetic scattering (blue), and the background (dashed black). The fit was convolved with the experimental resolution -- see the Supplementary Information for full details. At $\Q=(0, 0)$ magnetic excitations cannot be observed due to the very strong elastic specular scattering and the peak shape primarily reflects the experimental resolution function. The intensities are presented, normalized to the spectral weight in the 1-3~eV region containing the $dd$ excitations. The spectra at $\Q=(0, 0)$ have been divided by the factors written on the plot in order to make them visible on the same scale. Error bars indicate the magnitude of the statistical variations in the summed spectra.}
\label{Fig2}
\end{figure*}

RIXS spectra were measured on all samples from $\Q=(0,0)$ to $\Q=(0.4,0)$ (i.e.\ up to 80\% of the Brillouin zone boundary) and are plotted in Fig.~\ref{Fig2}. At $\Q=(0,0)$ strong specular elastic scattering dominates the spectrum. At higher \Q{} we observe a peak on the energy scale of 300~meV. For the undoped ($x=0$) case this peak corresponds to a magnon \cite{Braicovich2010,BraicovichPRB2010,Dean2012}, as observed previously by neutron scattering \cite{Coldea2001}. Examining the spectra at higher doping levels shows the central result of this paper: that this magnon smoothly broadens with doping and persists as a paramagnon right across the La$_{2-x}$Sr$_x$CuO$_4$ phase diagram and into the heavily overdoped metallic region. Understanding this observation and how it might relate to HTS is the main challenge posed by our data.

\begin{figure*}
\includegraphics{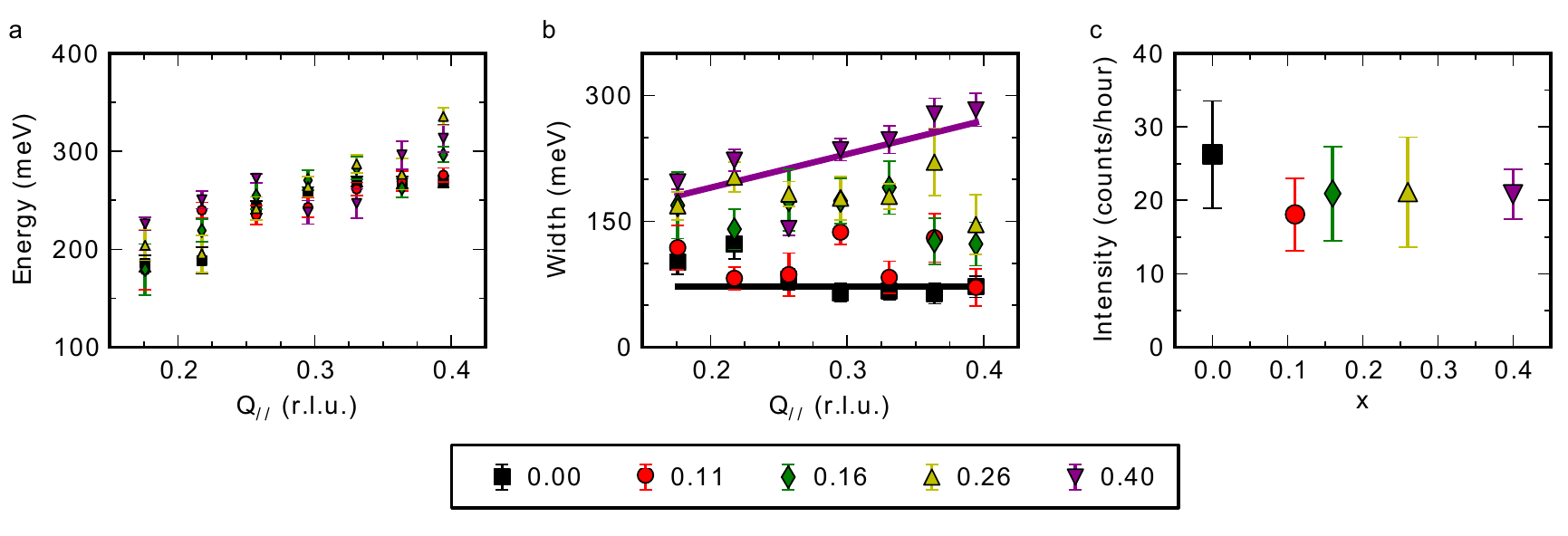} %
\caption{\textbf{Evolution with doping of the magnetic excitations in La$_{2-x}$Sr$_x$CuO$_4$} (a) Energy dispersion of the magnetic excitations along $\Q=(Q_{||},0)$, showing that the magnetic excitation energy does not change significantly with $x$. (b) The $Q_{||}$ dependence of the half width at half maximum (HWHM) of the magnetic excitations. The black and purple lines are guides to the eye for the $x=0$ and $x=0.40$ data respectively, emphasizing the overall increase of width with $x$. (c) The RIXS intensity of the magnetic excitations averaged over the measured \Q{} show that, within errors, the intensity of the magnetic excitations is conserved with increasing $x$. The error bars in panels (a) and (b) represent the uncertainty in the least-squares fitting routines, while in panel (c) these are combined with an uncertainty due to the instrumental instability (see the Supplementary Information for more details).}
\label{Fig3}
\end{figure*}

We fit the spectra with a resolution-limited Gaussian to account for the elastic scattering and an anti-symmetrized Lorentzian to account for the magnetic scattering, together with a smooth background (see the Supplementary Information for more details.) For the $x=0$ sample, the magnon is well defined in energy \cite{Coldea2001} and the peak broadening arises primarily from contributions from phonons and multimagnon excitations. This leads to a roughly $\Q$-independent broadening of the peak \cite{LeTacon2011}. Figure \ref{Fig3} summarizes the fitting results. Within the scatter of the data points, which is about $\sim 70$~meV (or $\sim$20\% of the zone boundary energy), the dispersion of the magnetic excitations in Fig.~\ref{Fig3}(a) is unchanged. In contrast, the width of the excitations in Fig.~\ref{Fig3}(b) increases dramatically with doping. This damping of the magnon with doping is likely due to the magnetic excitations coupling with, and decaying into, Stoner quasiparticles as low energy electronic states become available.

Measurements spanning the underdoped to slightly overdoped regions of the phase diagram in neodymium (NBCO) \cite{LeTacon2011} yttrium (YBCO) \cite{LeTacon2011} and bismuth (BSCCO) \cite{Dean2013a} based cuprates show comparable widths to our optimally doped La$_{1.84}$Sr$_{0.16}$CuO$_4$ sample. Our experiments go beyond optimal doping \cite{LeTaconPrivate} and show that the width continues to increase as more electronic states become available for scattering. The similar widths of optimally doped LSCO, YBCO and BSCCO are interesting in light of the fact that LSCO is often being regarded as more disordered than YBCO \cite{Alloul2009}. This may imply that the width is primarily determined by the hole concentration level rather than the structural disorder induced by doping.

We also examined the RIXS intensities as a function of doping. As discussed in the SI, instrumental instabilities can make measuring RIXS excitations in absolute units challenging \cite{BraicovichPRB2010}. For this reason, we present the \Q{}-averaged integrated intensity of the paramagnon in Fig.~\ref{Fig3}(c). Remarkably, the paramagnon is seen to have comparable RIXS intensity across the entire phase diagram, right into the heavily overdoped region.

These results heavily constrain any theoretical description of the magnetism in the cuprates. Motivated by the local moment physics in La$_2$CuO$_4$, and possible electronic phase separation in the underdoped cuprates, much theoretical work has described magnetism in the cuprates in terms of residual local moments which exist in the charge-poor region of the lattice \cite{Vojta2009}. The effect of doping is then primarily to reduce the number of local moments present. In the heavily overdoped region of the phase diagram, such phase separation seems unlikely to be energetically favourable as there is a large kinetic energy cost to localizing the electrons. Despite this, our data show that the local moment-based magnon excitations evolve smoothly and continuously across the phase diagram, into the region where renormalized itinerant quasiparticles have been suggested as explanations for the magnetic response of the cuprates \cite{Brinckmann2001}. In particular, our results demonstrate that the heavily overdoped region of the phase diagram retains magnetic correlations and is not a simple non-magnetic Fermi-liquid.

It is important to note that we report excitations from $(0.18,0)$ to $(0.40, 0)$. Inelastic neutron scattering experiments, on the other hand, are predominantly sensitive to the higher intensity excitations around $\Q_{\text{AFM}}=(0.5,0.5)$  where studies of heavily overdoped cuprates report a strong reduction in the energy and spectral weight of the magnetic excitations \cite{Wakimoto2007,Fujita2012NS}. In the region of \Q{}-space studied here, the intensity of the magnetic excitations are below the signal-to-noise ratio of current \Q{}-resolved neutron scattering experiments on doped cuprates. Thus RIXS allows us to measure magnetic excitations in a complementary region of the Brillouin zone.  While in N\'{e}el ordered La$_2$CuO$_4$ the energy dispersion is symmetric upon reflection in the antiferromagnetic Brillouin zone boundary, the same need not be true for $x\gtrsim0.03$. Furthermore, the intensities differ substantially depending on whether \Q{} is near $(0,0)$ or $\Q_{\text{AFM}}=(0.5,0.5)$. Taken together, the RIXS and neutron data imply that doping does not uniformly reduce the intensity of the magnetic excitations. Rather, the excitations around $(0.18\rightarrow0.40, 0)$ remain relatively unchanged in integrated intensity while the absolute intensity of the excitations around $\Q_{\text{AFM}}$ (which provide most of the total spectral weight) are strongly attenuated. The energy dispersion of the magnetic excitations along $(0.18\rightarrow0.40, 0)$ also remains constant as as function of doping, while the dispersion near $\Q_{\text{AFM}}$, is strongly renormalized to form the ``hour-glass'' feature \cite{Arai1999,Bourges2000, Tranquada2004, Hayden2004, Vignolle2007, Xu2009}.

Our results indicate that magnetic excitations evolve smoothly across the phase diagram. We do not see sharp changes in the high-energy magnetic fluctuations that one might expect if a hidden quantum critical point existed near optimal doping, as has been suggested. Our results seem to be consistent with the description of magnetic properties of the cuprates across the phase diagram based on the single-band two-dimensional Hubbard model. Numerical studies of this model \cite{scalapino2007} indeed qualitatively capture both the suppression of intensity around $\Q_{\text{AFM}}$ and the approximate conservation of intensity around $\Q = (0.18\rightarrow0.40, 0)$.

The cuprate phase diagram (see Fig.~\ref{Fig1}(a)) shows a dome-like dependence of $T_c$ as a function of $x$. On the underdoped ($x < 0.16$) side of the phase diagram the drop in $T_c$ and the eventual disappearance of HTS is likely due to the reduction in the density of mobile charge carriers. The reduction in $T_c$ on the overdoped ($x > 0.16$) side, in the face of the increase in the conductivity and the density of itinerant carriers, is likely driven by a reduction in the strength of the pairing interaction. Our results show that the demise of HTS in the overdoped cuprates is not due to changes in the high-energy magnetic excitations, as these remain approximately constant as a function of $x$. The change in $T_c$ must, therefore, be driven by other factors. These could include the influence of the low-energy magnetic excitations which are known to change dramatically in the overdoped cuprates. It is also conceivable that factors not captured within the simple Hubbard model may be at play.

\section*{Acknowledgments}
M.P.M.D. and J.P.H.\ are supported by the Center for Emergent Superconductivity, an Energy Frontier Research Center funded by the U.S.\ DOE, Office of Basic Energy Sciences. Work at Brookhaven National Laboratory was supported by the Office of Basic Energy Sciences, Division of Materials Science and Engineering, U.S. Department of Energy under Award No.\ DEAC02-98CH10886. This work was also partially supported by the Italian Ministry of Research MIUR (Grant No.\ PRIN-
20094W2LAY). The experiment was performed using the AXES instrument at ID08 at the European Synchrotron Radiation Facility. We acknowledge insightful, continuing discussions with Andrew James, Robert Konik and John Tranquada.

\section*{Methods}
\textbf{Sample preparation:} %
 The La$_{2-x}$Sr$_x$CuO$_4$ films were synthesized by atomic-layer-by-layer molecular beam epitaxy (ALL-MBE) \cite{Bozovic2001}. We used single-crystal LaSrAlO$_4$ substrates polished with the surface perpendicular to the $c$-axis, i.e.\ the [001] crystallographic direction. During film deposition, the substrates were kept at $T_s \approx 680 ^{\circ}$C under ozone partial pressure $p = 5\times10^{-6}$~Torr. The $x=0$, $x=0.16$ and $x=0.40$ films were 40 unit cells (53~nm) thick and the $x=0.11$ and $x=0.26$ films were 75 units cells (99 nm) thick. Reflection high-energy electron diffraction was monitored in real time to ensure that the films were atomically smooth and without any secondary-phase precipitates, which was verified \emph{ex situ} by atomic force microscopy. The undoped ($x=0$) and underdoped ($x=0.11$) films were annealed under vacuum ($2\times10^{-8}$~Torr) for one hour in order to drive out any excess (interstitial) oxygen. X-ray diffraction confirmed high crystalline order and absence of any secondary phases. To evaluate superconducting properties, magnetic susceptibility measurements were performed using the mutual inductance technique. Sharp superconducting transitions were observed in the films with $x = 0.11$ (underdoped, $T_c = 29$~K), 0.16 (optimally doped, $T_c = 39$~K) and 0.26 (overdoped, $T_c = 22$~K); in contrast, the undoped ($x=0$) and heavily overdoped ($x=0.40$) films showed no signs of superconductivity. More details are given in the Supplementary Information.

\textbf{RIXS measurements:} RIXS experiments were performed using the AXES instrument at the ID08 beamline at the European Synchrotron Radiation Facility. The spectra were typically collected for 1-2 hours, during which time the CCD camera was read out every 5 minutes. The incident x-ray energy was set to the peak in the measured Cu $L_3$ edge x-ray absorption spectrum. The x-rays were incident at angle $\theta_i$ with respect to the sample surface and scattered through $2\theta=130^{\circ}$ in the horizontal scattering plane (see the diagram in the supplementary information). We denote the in-plane scattering vector \Q{} in reciprocal lattice units (r.l.u.) using the 2D cuprate unit cell shown in Fig.~\ref{Fig1}(b) with $a=b=3.78$~\angstrom{}. To scan \Q{} we varied $\theta_i$ by rotating the sample about the vertical axis, thus varying the projection of \Q{} into the $a^*b^*$-plane, where $a^*$ and $b^*$ are the reciprocal lattice vectors. Data reported here used $\pi$ polarized x-rays, with high-\Q{} corresponding to x-rays that leave the sample at grazing angles. The combined resolution function of the monochromator and spectrometer is approximately Gaussian with a HWHM of 130~meV as determined by measuring the non-resonant elastic scattering from disordered carbon tape. All data were collected at $T = 25(5)$~K.

\section*{Author contributions}
Experiment: M.P.M.D., G.D., G.G., R.S.S., T.S., F.Y.-H., K.K., N.B.B.\ and L.B.
Data analysis and interpretation: M.P.M.D., J.P.H.\ and X.L.
Sample growth I.B.
Sample characterization: I.B., Y.-J.S.\ and J.S.
Project planning: M.P.M.D., J.P.H.\ and I.B.
Paper writing. M.P.M.D., J.P.H.\ and I.B.

\end{document}